\documentclass[aps,prd,twocolumn,10pt,superscriptaddress,showpacs,showkeys,floatfix,nofootinbib]
{revtex4-1}

\usepackage[intlimits]{amsmath}
\usepackage{amssymb}
\usepackage[utf8]{inputenc}
\usepackage{multirow}
\usepackage{amsfonts}
\usepackage{amssymb}
\usepackage{setspace}
\usepackage{graphicx}
\usepackage{textcomp}         
\usepackage{array}
\usepackage{slashed}
\usepackage[usenames,table]{xcolor}
\usepackage[colorlinks=true,linkcolor=blue,citecolor=blue,urlcolor=blue]{hyperref} 
\usepackage[sort&compress]{natbib}

\begin{document}
%
%******************************************************************************************
% Title
%******************************************************************************************
%
\title{On the low-energy spectrum of an SU(2) gauge theory with dynamical fermions}

\author{Milan Vujinovic}
\email{milan.vujinovic@ifsc.usp.br}

\affiliation{Institute of Physics, NAWI Graz, University of Graz,  
 Universit\"atsplatz 5, 8010 Graz, Austria}

\affiliation{Sao Carlos Institute of Physics, University of Sao Paulo, Brazil}

\author{Reinhard Alkofer}
\email{reinhard.alkofer@uni-graz.at}

\affiliation{Institute of Physics, NAWI Graz, University of Graz,  
 Universit\"atsplatz 5, 8010 Graz, Austria}
%
%******************************************************************************************
% Abstract
%******************************************************************************************
%
\begin{abstract}

The spectrum of light bound states in an SU(2) gauge theory 
with two flavors of fundamentally charged fermions is investigated 
by solving the Bethe-Salpeter equations in the respective channels within a 3PI-type
({\it i.e.}, beyond rainbow-ladder) truncation including self-consistently a 
correspondingly truncated fermion--gauge-boson vertex. Remarkable differences 
with respect to the meson spectrum of an SU(3) gauge theory are found 
although in our approach these are not as pronounced  as indicated by 
some recent respective investigations within lattice gauge field theory.

\end{abstract}

%\pacs{12345}

%\keywords{something here}

\maketitle
%
%
%********************************************************************************************
% Introduction
%********************************************************************************************
%
%
\section{Introduction}\label{sec:introduction}

Understanding the physics of bound states in generic gauge field theories is for several
reasons of interest. Over the last decades great efforts have been undertaken to describe
hadrons as bound states of quarks and glue within QCD. Certain types of 
Beyond-Standard-Model (BSM) approaches as composite Higgs models 
\cite{Kaplan:1983sm,Agashe:2004rs}
and technicolor theories \cite{Farhi:1980xs,Andersen:2011yj}
also require to understand the physical spectrum of quantum field theories, including 
those of the strongly-interacting type. Last but not least, dark matter might occur in 
the form of bound states within a hidden strongly-interacting sector, {\it cf.}, 
the so-called SIMP scenario \cite{Hochberg:2014dra}.

Only recently the necessary tools for studying bound states in strongly-interacting 
quantum field theories have been developed. On the one hand, in lattice gauge theories 
(hadron) spectroscopy has proven to be a much more  complicated task than previously 
expected, see, {\it e.g.}, ref.\ \cite{Lang:2015ljt} and references therein. 
On the other hand, due to the rich and quite often complicated structure 
of highly relativistic bound states of elementary constituents with spin, studies of 
hadrons within functional methods have been restricted to generalized rainbow-ladder (RL)
truncations of the Bethe-Salpeter equation until some years ago, see, {\it e.g.}, 
ref.\ \cite{Eichmann:2016yit} and references therein, and even 
nowadays most of such investigations of bound-state properties like, {\it e.g.}, form 
factors and decays, still rely on RL-type approximations together with a phenomenologically
adapted momentum dependence of the constituents' interactions.

The description of relativistic bound states from Quantum Field Theory dates back to 
the seminal papers by Bethe and Salpeter \cite{Salpeter:1951sz}.
Whereas their treatment of the deuteron 
was based on an expansion of the kernel, modern functional methods emphasise the 
importance of symmetries, for reviews on either the framework of Dyson-Schwinger 
\cite{Dyson:1949ha,Schwinger:1951ex} and Bethe-Salpeter equations \cite{Salpeter:1951sz}
or the functional renormalization group \cite{Wetterich:1992yh,Morris:1993qb}, see 
\cite{Roberts:1994dr, Alkofer:2000wg, Fischer:2006ub, Cloet:2013jya,Berges:2000ew,
Pawlowski:2005xe}. 
Within the Dyson-Schwinger--Bethe-Salpeter framework the simplest symmetry-preserving truncation scheme is given by keeping the sum of all rainbow diagrams in the 
one-particle self-energy, ({\it i.e.}, in the two-point function) and all ladder 
diagrams in the four-point function. This RL truncation is the simplest scheme which 
obeys the constraints from the axial-vector Ward-Takahashi identity. This is an 
important feature in QCD as it guarantees the Goldstone boson nature of the pion. 
However, despite its considerable successes in the phenomenological studies of mesons and 
baryons, this and related truncation schemes have some serious practical and conceptual
limitations, see, {\it e.g.}, 
\cite{Eichmann:2016yit,Maris:1997hd,
Maris:1997tm,Maris:1999nt,Krassnigg:2009gd,Krassnigg:2009zh,Blank:2010pa,Krassnigg:2010mh,
Blank:2011ha,Roberts:2011cf,Fischer:2014xha,Hilger:2014nma,Fischer:2014cfa, 
Hilger:2015hka,Eichmann:2009qa,
Nicmorus:2010sd,SanchisAlepuz:2011aa,Eichmann:2011vu,Sanchis-Alepuz:2013iia, Sanchis-Alepuz:2015fcg,Sanchis-Alepuz:2017jjd} 
for the successes and shortcomimgs of RL-type truncations for mesons and baryons.

Therefore one of the long-standing goals within functional methods is to establish more
sophisticated truncation schemes that can be systematically applied to reliably 
calculate bound-state properties. This task can be approached in two different ways: 
bottom-up or top-down. While the former uses phenomenological input in order to 
construct models and determine their parameters, the latter requires a robust 
theoretical foundation upon which to build. Consequently, there is a rich and 
diverse history regarding truncations of relativistic bound state equations.%\footnote{
(NB: BRL truncations can be roughly categorized as diagrammatic, see, {\it e.g.}, 
\cite{Munczek:1994zz,Bender:1996bb,Bender:2002as,Watson:2004kd,Watson:2004jq,Bhagwat:2004hn,
Matevosyan:2006bk,Fischer:2009jm,Windisch:2012de,Gomez-Rocha:2014vsa,Williams:2014iea, Vujinovic:2014ioa,Sanchis-Alepuz:2015tha, Sanchis-Alepuz:2015qra, Williams:2015cvx},
and non-diagrammatic approaches, see, {\it e.g.}, 
\cite{Fischer:2005en,Fischer:2007ze,Fischer:2008sp,Fischer:2008wy,
Chang:2009zb,Chang:2011ei,Heupel:2014ina}.)
Recent investigations have proven that it is essential to solve, at least, for the 
three-point vertices of the elementary constituents explicitly 
in an (at least approximately) self-consistent procedure. 
In ref.\ \cite{Williams:2015cvx} light mesons have been investigated in the so far most
sophisticated truncation scheme to QCD within the Dyson-Schwinger--Bethe-Salpeter 
framework. Based on the three-particle irreducible effective action quark-loop 
contributions to the gluon propagator and three-gluon vertex have been taken into account. 
The resulting fully coupled system of Dyson-Schwinger equations for two- and three-point 
functions have been solved self-consistently. The symmetry-preserving quark-antiquark kernel 
of the Bethe-Salpeter equation for mesons has been derived and time-like properties of 
bound-states have been obtained by analytic continuation of Euclidean momenta.

Related studies of QCD have been performed recently with functional renormalization group 
techniques, see ref.\ \cite{Cyrol:2017ewj} and references therein. 
Employing a vertex expansion 
scheme based on gauge-invariant operators a quantitative analysis of chiral symmetry 
breaking has been performed, and the feasibility of dynamical hadronization has been 
demonstrated. The resulting quark propagator, quark-gluon vertex (including its full tensor 
structure and momentum dependence), and some properties of the four-fermion scattering 
kernels have been calculated. However, the analytic continuation necessary to discuss bound 
states could not be achieved yet. This shortcoming in mainly due to the lack of suitable 
regulator functions which are a defining element of the functional renormalization group 
equations. Recent progress on this issue \cite{Paris-Lopez:2018vjc} 
demonstrates the technical nature of this 
limitation and makes evident that it will be overcome in the near future. 

In the last years SU(2) gauge theories have been studied mainly for two reasons. 
Two-color gauge theories with an even number of fermion flavors $N_f$ have been of 
interest to lattice practitioners because  
Monte-Carlo lattice simulations of them at non-vanishing chemical potential
are not hindered by the sign problem, and one can gather information about the 
phase diagram of the corresponding strongly-interacting matter on the lattice
\cite{Hands:1999md,Aloisio:2000rb,Hands:2000ei,Kogut:2001na,Muroya:2002ry,
Hands:2006ve,Cotter:2012mb,Bahrampour:2016qgw}. 
This then also initiated related studies with
functional methods \cite{Vujinovic:2014ioa,Khan:2015puu,Contant:2017gtz}.
Recently, an SU(2) gauge theory with two fundamentally charged Dirac fermions 
has been studied on the lattice 
\cite{Arthur:2016dir,Arthur:2016ozw,Drach:2017jsh,Drach:2017btk,Lee:2017uvl}
because it provides the simplest field theoretical realization of a unified theory of
a composite Goldstone boson Higgs and technicolor \cite{Cacciapaglia:2014uja}. 
Therefore such a theory might serve as a template for aspects of dynamical 
electroweak symmetry breaking as well as the SIMP scenario for Dark Matter.

In the here presented investigation light fermion-anti\-fermion (“mesons”) 
and fermion-fermion 
(bosonic two-color “baryons”) bound states are studied in an SU(2) gauge theory
with two fundamentally charged fermions within a beyond-rainbow-ladder (BRL)
truncation to the respective Dyson-Schwinger (DS) and Bethe-Salpeter (BS) equations.
A certain focus is given on an analysis of the impact of various diagrammatic 
contributions to the fermion--gauge-boson--vertex function on bound-state observables. 
Hereby suitable model input for the Yang-Mills two- and three-point functions is used 
to evaluate the fermion--gauge-boson vertex in a semi-self-consistent way. While there are 
similar calculations available \cite{Vujinovic:2014ioa, Sanchis-Alepuz:2015qra}, 
the present study improves upon these %by including the so-called Abelian loops 
in the truncation for the quark-gluon-vertex DS equation, for details see 
sect.~\ref{sec:quark_gluon}.

The paper is organized as follows: In sect.~\ref{sec:formal} the employed bound-state
equations as well as the determination of the necessary input are provided and discussed, 
especially also with respect to preserving chiral symmetry. 
In sect.~\ref{sec:quark_gluon} the used truncations for the equation of the 
fermion--gauge-boson vertex are introduced. These constitute the main element of the 
BRL truncation utilized in the following, and therefore it is discussed 
how different elements of the coupled system of DS equations for the fermion--gauge-boson
vertex and the fermion propagator influence this fundamental vertex function. 
In sect.~\ref{sec:kernel} a symmetry-preserving kernel of the bound state 
equation is presented, and 
in sect.~\ref{sec:mesons} the spectrum of light fermion-antifermion and
fermion-fermion bound states  in a $SU(2)$ gauge theory with two fundamentally charged
fermions is presented. In sect.~\ref{sec:conclude} our conclusions are provided.
Some technical issues related to solving for the fermion--gauge-boson vertex are
defered to Appendix~\ref{app:quark_gluon}. 
 
%
%
%********************************************************************************************
% Formalism
%********************************************************************************************
%
%
\section{The bound-state equation}\label{sec:formal}

\subsection{Constraining the kernel}

In an SU(2) gauge theory mesonic-type and baryonic-type bound states are both two-body 
bound states and bosons; their respective channels are related by the so-called 
Pauli-G\"ursey symmetry and are therefore degenerate. In the following it is therefore
completely sufficient to focus on the fermion-antifermion bound states to understand
the low-lying spectrum. %\footnote
{Nevertheless, degeneracy factors of the multiplets 
have to take into the account the existence of fermion-fermion (baryonic-type) bound states.}

Understanding the quantum numbers of possible Goldstone bosons which might appear in the
chiral limit will provide us some guiding principle when choosing a truncation 
to the bound state equation, see below the discussion of the the axial-vector 
Ward-Takahashi identity (axWTI).
The pseudo-reality of the fundamental representation of the group SU(2) implies also 
that the flavor symmetry is upgraded to SU(4). Dynamical chiral symmetry breaking leaves
a Sp(4) (locally isomorphic to SO(5)) intact, and therefore one expects to have 
in the chiral limit five Goldstone bosons: Three of them are fermion-antifermion bound 
states (similarly to the  pions in QCD), and two of them are of the baryon-, 
respectively, diquark-type. Note that in the latter case the Pauli principle reduces 
the number of states, {\it cf.\/} for example the discussion in
ref.~\cite{Alkofer:1995mv} and references therein. 
Note also, that a meson-type state with $J^{P}$ quantum numbers will have the same 
mass as a diquark-type state with equal total spin and opposite parity ($J^{-P}$). 
Thus, with the meson spectrum obtained, one can immediately extend the results and conclusions to the two-fermion states as well.

Using this setup as a model
for electroweak symmetry breaking the ``direction'' of the 
symmetry breaking with respect to Standard Model determines the nature of the model:
For a vanishing angle one obtains a composite Higgs model where four of the five
Goldstone bosons provide the Higgs doublet, and the fifth is neutral under Standard Model
charges. At the maximal angle one obtains a technicolor theory. Three of the Goldstone
bosons enter as members of BRST quartets, and thus the longitudinal components of the 
$W$ and $Z$ bosons become physical. The Higgs is then the lightest scalar bound state,
and the remaining two Goldstone bosons can be considered as Dark Matter candidates. 
In case the angle is determined dynamically to be neither vanishing nor maximal, 
the Goldstone boson Higgs mixes with the 
technicolor scalar bound state, and the lighter of the two scalars will be identified 
with the physical Higgs. Needless to say that then the spectrum of this theory will 
be more complicated to understand than in the two extreme cases. Last but not least, 
the back-coupling to Standard Model particles will lead to large corrections (see 
e.\,g.~\cite{Foadi:2012bb}),
and therefore quite some elaborate studies are required before one can
judge the usefulness of such a theory as a Beyond-the-Standard-Model scenario.

\begin{figure}[t]
\begin{center}\includegraphics[height=2.55cm]{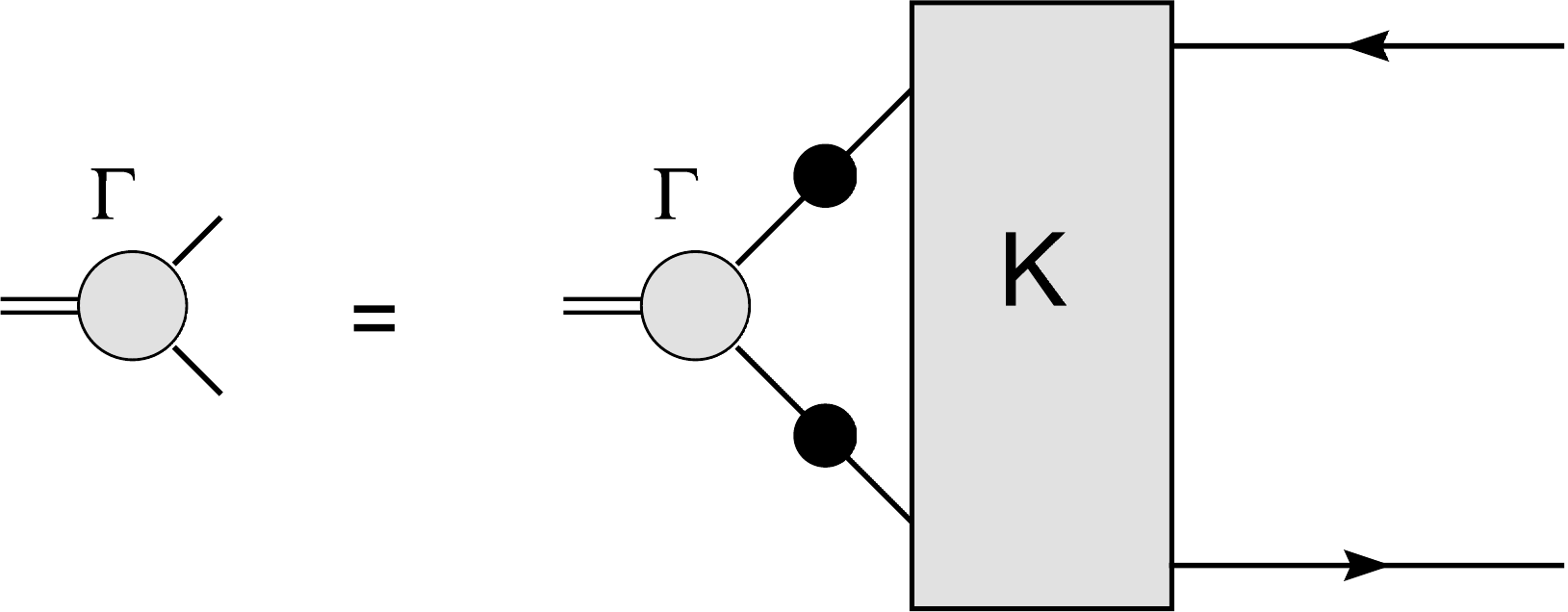}
\caption{The meson Bethe--Salpeter equation.}\label{fig:mesonbse}
\end{center}
\end{figure}

The two-body BS equation, depicted in fig.~\ref{fig:mesonbse},
for the amplitude $\Gamma_M(p,P)$ takes generically the form (see,
{\it e.g.}, \cite{Alkofer:2000wg}) 
\begin{equation}\label{eqn:mesonbse}
\left[\Gamma_M(p, P)\right]_{ij} = \int_k\left[K(p,k,P)\right]_{ik;lj} \left[\chi_M(k,P)\right]_{kl} \;.
\end{equation}
Hereby, $P$ denotes the total four-momentum of the bound state, and $p$ the relative momentum
between the constituents. It is implicitely understood that the amplitude is projected 
onto an eigenstate of the Pauli-Lubanski vector and on parity and charge conjugation 
eigenstates, {\it i.e.}, $J^{PC}$ are good quantum numbers. The abbreviation 
$\int_k$ stands for $\int d^4k/(2\pi)^4$, and $K(p,k,P)$ denotes the interaction kernel.
$\chi_M(k,P)$ is the so-called BS wave-function which is related to amplitude via the
relation 
\begin{equation}
\chi_M(k,P)=S(k_+)\Gamma_M(k,P)S(k_-)
\end{equation}
with $S(k_{\pm})$ being the fermion propagator. Although there is some freedom in the momentum assignement we choose the momenta $k_\pm$ to be $k_\pm = k \pm P/2$ which is 
the optimal choice when solving eq.~\eqref{eqn:mesonbse} numerically. 

\begin{figure}[b]
\begin{center}\includegraphics[height=1.01cm]{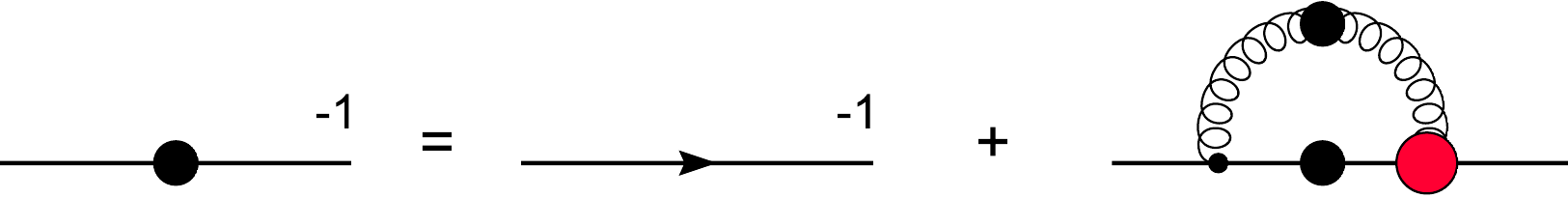}
\caption{The DS equation 
for the fermion propagator. Straight lines are quarks, wiggly ones gluons. 
Filled circles denote full propagators and vertices.}\label{fig:quarkdse}
\end{center}
\end{figure}

An essential ingredient needed for the evaluation of the BS equations is the fermion
propagator $S(k)$. In covariant gauges, it can be decomposed as 
\begin{align}\label{eqn:quark}
S^{-1}(p) = Z_f^{-1}(p^2)\left[ i\slashed{p} + M(p^2)\right] \;,
\end{align}
where $Z_f(p^2)$ is the respective fermion wavefunction renormalization,  
and $M(p^2)$ is a dynamically generated mass function. 
At tree-level, the above expression simplifies to 
$S^{-1}_0(p) = i\slashed{p} + Z_m m$, with
$Z_m$ being the fermion mass renormalization constant. 
The fermion two-point function satisfies its own Dyson-Schwinger equation, 
given by (see also Fig.~\ref{fig:quarkdse})
\begin{align}\label{eqn:quarkdse}
S^{-1}(p) & = Z_2S_0^{-1}(p) \\
          & + g^2 Z_{1f}C_f\int_k \gamma^\mu S(k+p)\Gamma^\nu(k+p,p)D_{\mu\nu}(k)\;.
\nonumber
\end{align}
The functions
$D_{\mu\nu}(k)$ and $\Gamma^\nu(k+p,p)$ are, respectively, the full gauge-boson propagator
and the fermion--gauge-boson vertex. 
$C_f$ is the gauge group's Casimir invariant in the fundamental representation, with 
$C_f = 3/4$ in an $SU(2)$ gauge theory. $Z_2$ and $Z_{1f}$ are the renormalization
constants for the fermion field and the fermion--gauge-boson three-point interaction, 
respectively. In the next section we provide details
on how the various renormalization factors are obtained.

The four-point kernel $K(p,k,P)$ of eq.~\eqref{eqn:mesonbse} subsumes an infinity 
of processes through which a fermion and an anti-fermion can interact. 
Obviously, any practical consideration of 
the BS equation requires the interaction kernel to be truncated. 
In the studies of light-light and heavy-light mesons, an important guideline for these 
truncations has been and is provided by the axial-vector 
Ward-Takahashi identity (axWTI), which connects the respective four-point function 
$K$ to the quark self-energy $\Sigma(k)$ of eq.~\eqref{eqn:quarkdse}:
\begin{align}\label{eqn:axWTI}
[\Sigma(p_+)\gamma_5 & + \gamma_5\Sigma(p_-)]_{ij} = \\
                    &\int_k \left[K(p,k,P)\right]_{ik;lj} \left[\Sigma(k_+)\gamma_5  + \gamma_5\Sigma(k_-)\right]_{kl}\;.\nonumber
\end{align}

\begin{figure}[b]
\begin{center}\includegraphics[height=1.93cm]{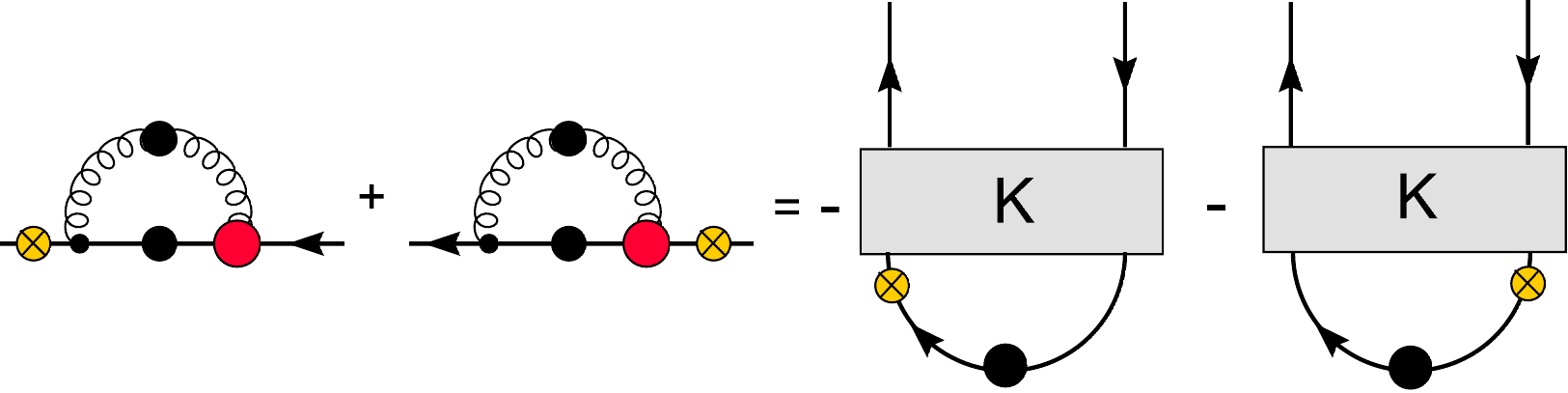}
\caption{A diagrammatic form of the axWTI for flavour non-singlet mesons. Yellow blob stands for the $\gamma_5$ matrix.}\label{fig:axwti}
\end{center}
\end{figure}

A diagrammatic representation of the axWTI for flavor non-singlet mesons is given in 
fig.~\ref{fig:axwti}. If a particular truncation for the quark DS and meson BS  
equation satisfies this  identity, the special status of light pseudoscalar mesons 
as (pseudo-)  Nambu-Goldstone bosons will remain intact, 
and the masslessness of the pseudoscalar ground states in the exact chiral limit 
is guaranteed. As it is expected that chiral symmetry and its breaking patterns play
an important role also in technicolor and/or composite Higgs models we require in 
the following the axWTI-induced relation between the kernels of the BS and the DS equations.

\begin{figure}[b]
\begin{center}
\includegraphics[width = 0.3\textwidth]{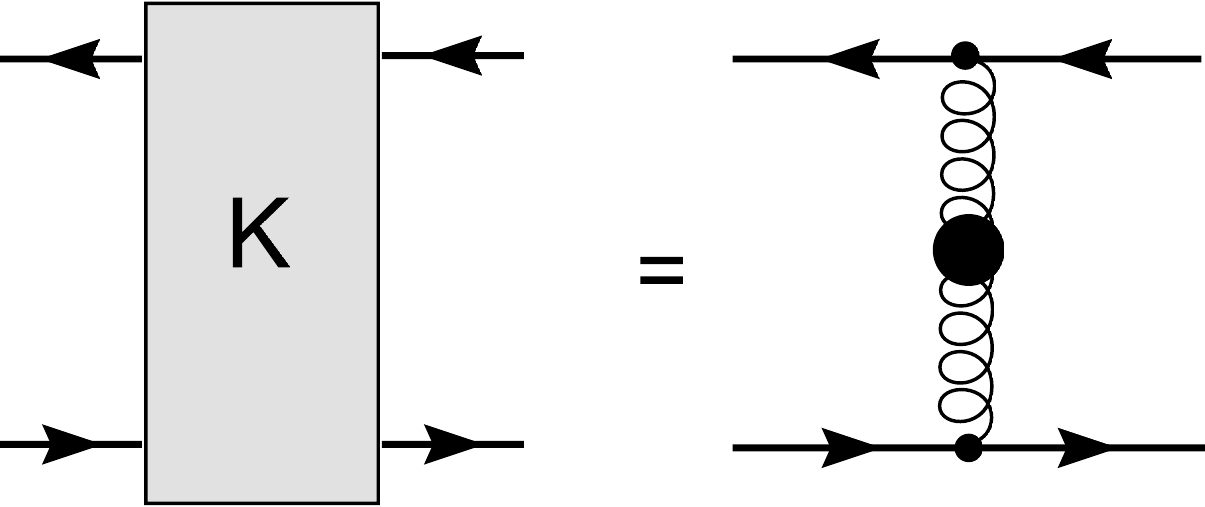}
\caption{The ladder truncation of the meson BSE interaction kernel.}\label{fig:ladder}
\end{center}
\end{figure}

One  way to obtain an axWTI-preserving kernel $K$ from an approximated fermion DS
equation is to require the kernel $K$ in coordinate-space to be given by the following 
functional derivative of the fermion's self-energy $\Sigma$ with respect to 
the the fermion's propagator (also in coordinate-space)
\begin{equation}\label{eqn:cutting}
K(x_1,x_2,x_3,x_4) = -\frac{\delta \Sigma(x_1,x_2)}{\delta S(x_3,x_4)} \;.
\end{equation}
In a diagrammatic language, the operation of eq.~\eqref{eqn:cutting} corresponds to
``cutting'' all internal fermion lines in the fermion propagator DS
equation to generate the kernel of the BS equation \cite{Munczek:1994zz, Bender:1996bb}. 
An illustration is provided by the simplest non-trivial scheme which obeys the 
axWTI, the RL truncation.
Starting by replacing the fully dressed fermion--gauge boson vertex in 
eq.~\eqref{eqn:quarkdse} with its tree-level counterpart, possibly multiplied by a 
function of the gauge-boson's momentum squared, $\lambda(k^2)$, {\it i.e.},
\begin{equation}\label{eqn:rainbow}
\Gamma^\nu(k+p,p)\rightarrow \lambda(k^2)\gamma^\nu\; ,
\end{equation}
one applies the above described cutting technique. It is then straightforward to 
derive  that the corresponding symmetry-preserving BS kernel is given by an exchange 
of a single dressed gauge boson which is shown in fig.~\ref{fig:ladder}
and which consitutes one rung of the ``ladder''
generated by iteration of the kernel within the BS equation. 
In most of its hadron physics applications, the function $\lambda(k^2)$ of 
eq.~\eqref{eqn:rainbow} was 
combined with the non-perturbative dressing of the gluon propagator into a single 
effective interaction, and the model parameters are chosen such 
that some hadronic observables ({\it e.g.} the pion decay constant $f_\pi$ and the
$\rho$ vector meson mass $m_\rho$) are correctly reproduced. 

As mentioned in the introduction the BS approach in RL truncation has enjoyed considerable
successes in the phenomenological studies of hadrons, and it is still widely used today.
Of course, a major part of this success is related to choosing the interaction model 
parameters by fitting to a few hadronic observables. To which extent the function 
$\lambda(k^2)$  reflects properties of the quark-gluon vertex stays elusive. 
Since the whole formalism in the RL truncation is reflecting basically only the
properties of the calculated quark propagator, it is 
in general not possible to disentangle the various physical processes and 
interactions (like, {\it e.g.}, pion cloud effects) which contribute to measurable 
quantities \cite{Williams:2009wx, Sanchis-Alepuz:2014wea}, or to assess the influence 
of gauge degrees of freedom. Additionally, it is virtually impossible to adjust the 
RL framework to strongly-interacting theories different  from QCD, since there is no 
obvious and meaningful way to re-parameterize the effective interaction(s) to account 
for changes in the principal vertex functions of the model, see, {\it e.g.},
\cite{Vujinovic:2014ioa} as an example. Furthermore, in a solution for the coupled 
DS equations for the QCD propagators with a rainbow truncation for the quark DS 
equation the onset of the conformal window occurs for a too small number of quark 
flavors \cite{Hopfer:2014zna}. This can be related to overestimating the quark loop
in the gluon DS equation, and thus can only be improved if more than the tree-level
structures of the quark-gluon vertex are taken into account. This implies then 
for the current investigation to use a more general framework including especially
solving, at least approximately, for the fermion--gauge-boson three-point vertex. 

%
%
%**********************************************************************
% Quark-gluon vertex
%**********************************************************************
%
%
\subsection{The fermion--gauge-boson three-point vertex}\label{sec:quark_gluon}

The three-point fermion--antifermion--gauge-boson vertex possesses in covariant 
gauges in general twelve tensor components. In Landau gauge only those eight 
components are needed which are purely transverse to the gauge-boson's 
momentum. Therefore, and also because other needed correlation functions
are best known in this gauge, the investigation reported herein 
is done within the Landau gauge.

\begin{figure}[b]
\begin{center}
\includegraphics[width = 0.48\textwidth]{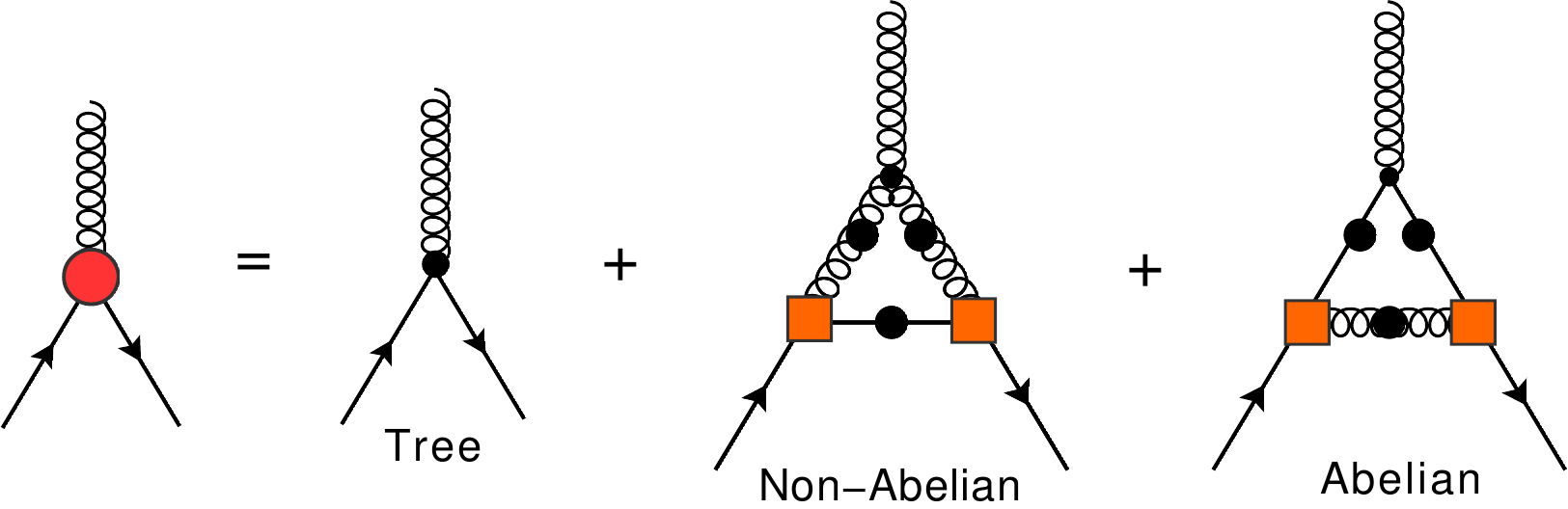}
\caption{The truncated fermion--gauge-boson three-point vertex DS  equation, 
in ``1PI'' formulation. The internal fermion--gauge-boson
vertices (orange squares) are modeled,
see Appendix~\ref{app:quark_gluon} for details.
 We also consider the ``3PI'' approximation, with all internal vertices dressed.}
\label{fig:qg_dse}
\end{center}
\end{figure}

The DS equation for the three-point fermion--gauge-boson vertex 
equation can be written in two different forms. These as well as the 
corresponding derivations may be, {\it e.g.}, found in ref.\ \cite{Alkofer:2008tt}, 
Chapter 2 of ref.\ \cite{Windisch:2014lce} or Chapter 7 of ref.\ \cite{Hopfer:2014szm}.
For the sake of brevity we focus immediately on the truncated form
shown in fig.~\ref{fig:qg_dse}. The truncation (see refs.\ 
\cite{Alkofer:2008tt,Williams:2014iea,Williams:2015cvx,Windisch:2014lce,Hopfer:2014szm} 
for its justification) consists of considering  the one-loop contributions 
which contain primitively divergent vertices. These then retains two diagrams on the 
r.h.s.\ of the truncated DS equation which are usually labeled as Non-Abelian and
Abelian diagram. Retaining one or both  of these two contributions is the usual
approximation when treating the quark-gluon vertex function in functional or 
semi-perturbative approaches, see, {\it e.g.}, refs.\
\cite{LlanesEstrada:2004jz,Alkofer:2008tt,Windisch:2012de,Hopfer:2013np,
Windisch:2014lce,Hopfer:2014szm,Williams:2014iea,Williams:2015cvx,Williams:2009wx,
Pelaez:2015tba,Fu:2015tdu} (but also refs.\ 
\cite{Rojas:2013tza, Aguilar:2014lha, Mitter:2014wpa,Oliveira:2018fkj} 
for significantly different continuum formulations).

As will become evident below it is not necessary to treat this three-point vertex 
function in a fully self-consistent way: Instead of back-coupling the full vertex function 
(the red blob in fig.~\ref{fig:qg_dse}) into its DS equation  a projected
version is used for the internal vertices (denoted by orange squares in 
fig.~\ref{fig:qg_dse}). As this point is of a completely technical nature all details
of this procedure are described in Appendix~\ref{app:quark_gluon}. It is sufficient
to note for the following that for the full vertex
the complete eight transverse components are used. For the purpose of solving its DS 
equation, however, we project it onto the respective tree-level component, 
thereby calculating an effective dressing $\lambda(k^2)$, and in turn we employ this on the 
r.h.s.\ of the vertex DS equation.

In the context of bound state studies the above-mentioned approximation provides a
significant and almost necessary technical simplification. The  most 
important reason for this approximation relates to the implementation of the cutting 
procedure for the construction of a symmetry-preserving BS kernel. If one were to employ the 
vertex in a fully self-consistent way one would have to take into account its 
implicit fermion propagator dependence, and the functional derivative in 
eq.~\eqref{eqn:cutting} would produce some very complicated terms in the bound 
state equation as, {\it e.g.}, a five-point Green function with four fermion 
and one gauge boson leg. While it is possible to obtain a solvable BS kernel 
with a self-consistent treatment of the fermion--gauge-boson vertex,
see, {\it e.g.}, refs.\ \cite{Sanchis-Alepuz:2015tha,Williams:2015cvx}, 
so far no one has tackled the challenge of solving the bound-state equation with the 
Abelian loops included in a self-consistent manner. Our calculation can thus be seen as an 
intermediate step towards a more complete treatment: Additional diagrams 
are included, both in the vertex equation  and the BS kernel, but the 
evaluation of the fermion--gauge-boson vertex  itself is considerably simplified
without loosing, at least partially, the back-coupling effect of the vertex on itself.

Besides the one-particle irreducible (1PI) equation, depicted in 
fig.~\ref{fig:qg_dse}, we also consider a form derived from 
the three-particle irreducible (3PI) formalism 
\cite{Berges:2004pu}, {\it cf.\/} also refs.\ 
\cite{Windisch:2014lce,Hopfer:2014szm,Hopfer:2013np,Williams:2015cvx}, 
with all the internal vertices dressed. The notation 1PI/3PI 
should hereby not be understood in a strict manner but more as convenient 
labels for the presentation of the results, mainly because  1PI or 3PI formulations 
would entail a self-consistent evaluation of all vertex functions. 
Note that additional vertex dressings in the 3PI approach can effectively 
be seen as a partial  inclusion of the disregarded two-loop terms. 
In addition, they also provide an estimate of the truncation errors. For the non-Abelian 
diagram in the 3PI framework also the fully dressed three--gauge-boson vertex is required.

%
%
%**********************************************************************
% Yang-Mills
%**********************************************************************
%
%
\subsection{Gauge boson correlation functions}

\begin{figure}[b]
\begin{center}
\includegraphics[width = 0.48\textwidth]{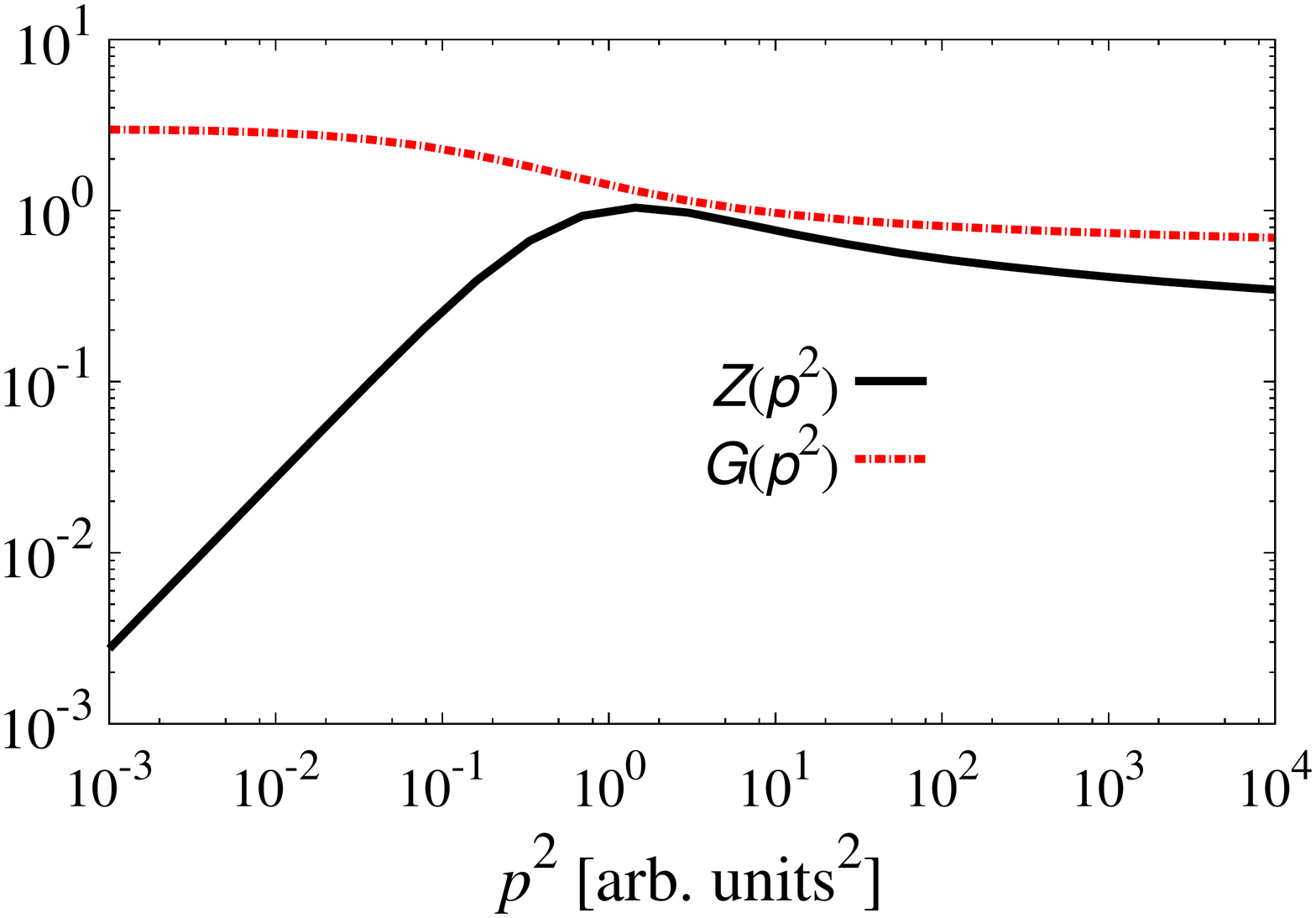}
\caption{Model dressing functions for ghost (\textit{G}) and gauge boson (\textit{Z}) 
propagators as function of the square of an arbitrary momentum scale.
The connection to physical scales is established in section
\ref{sec:mesons}.}\label{fig:yang_mills}
\end{center}
\end{figure}

As became evident in the discussion above the gauge-boson propagator and the  
three--gauge-boson vertex are needed as input. 
For determining the latter one also the ghost propagator will serve as input.
In Landau gauge the gluon and ghost propagators, $D_{\mu\nu}(k)$ and $D_G(k)$, 
respectively, are of the form
\begin{equation}\label{eqn:ghost_gluon}
D_{\mu\nu}(k) = T^{(k)}_{\mu\nu}\frac{Z(k^2)}{k^2}, ~~ D_G(k) = -\frac{G(k^2)}{k^2}\;,
\end{equation}
with $T^{(k)}_{\mu\nu} = \delta_{\mu\nu} - k_\mu k_\nu/k^2$ being the transverse projector
with respect to momentum $k$. The dressing functions $Z(k^2)$ and $G(k^2)$ can be 
determined from their respective DS equations. A compilation of results for
Yang-Mills correlation functions at different levels of truncations can be found 
in ref.\ \cite{Huber:2018ned}. 
An example for those dressing functions (which are then used also in the following) 
are shown in fig.~\ref{fig:yang_mills}.
Details of their calculation are given in ref.\ \cite{Eichmann:2014xya}. 
Obtaining the Yang-Mills propagators from their DS equations has the benefit of 
providing not only the required dressing functions but also other essential input 
such as renormalization constants. Hereby, $\tilde{Z}_3$ and $Z_3$ being, respectively, 
the ghost and gauge boson renormalization constants are used to determine
the corresponding renormalization constants for other Greens functions 
via Slavnov-Taylor identities. We will return to this point below when we 
will discuss the numerical method for the coupled system of DS
equations for the fermion propagator and the fermion--gauge-boson vertex.

\begin{figure}[b]
\begin{center}
\includegraphics[width = 0.47\textwidth]{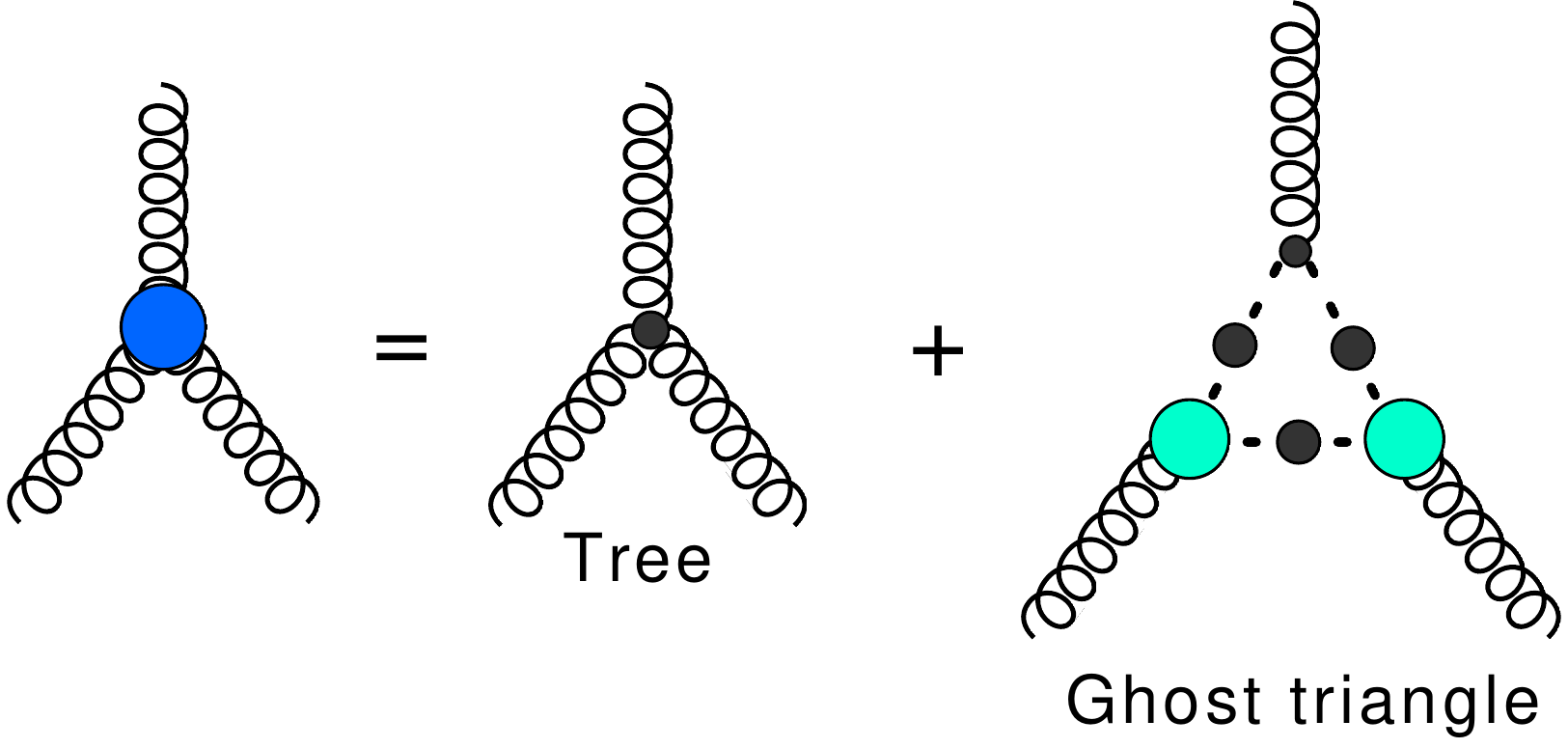}
\caption{A truncated DS equation for the three--gauge-boson vertex. 
The full ghost--gauge-boson vertices are approximated by bare ones.}\label{fig:3g_dse}
\end{center}
\end{figure}

\begin{figure}[t]
\begin{center}
\includegraphics[width = 0.47\textwidth]{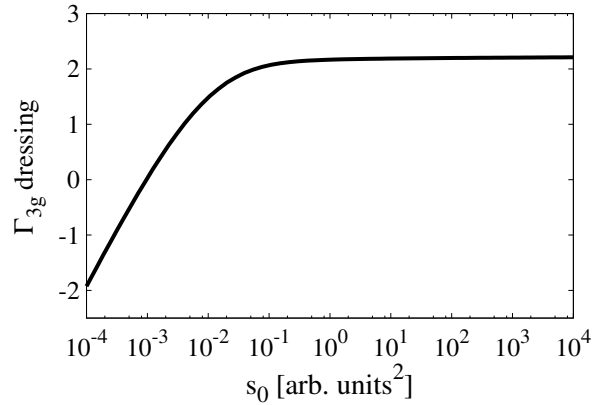}
\caption{Dressing for the three--gauge-boson vertex as a function of the  
momentum variable $s_0 = (1/6)\cdot(p_1^2 + p_2^2 + p_3^2)$ in arbitrary units. 
Connection to physical scales is established in section
\ref{sec:mesons}.}\label{fig:3g_model}
\end{center}
\end{figure}

For the three--gauge-boson correlation function we use the truncation depicted
in fig.~\ref{fig:3g_dse}. First of all, based on the
results of ref.\ \cite{Eichmann:2014xya} it is a fair approximation to keep for the
gauge group and Lorentz tensor structure only the tree-level ones. Hereby the
tree-level three--gauge-boson vertex is denoted
 by  $\Gamma^{(0)}_{\mu\nu\rho}(p_1,p_2,p_3)$. Second, the momentum
dependence of the multiplying function can be quite well represented by the form 
\begin{equation}\label{eqn:three_gluon}
\Gamma_{\mu\nu\rho}(p_1,p_2,p_3) = \mathcal{A}(s_0)\cdot\Gamma^{(0)}_{\mu\nu\rho}(p_1,p_2,p_3)\;
\end{equation}
where the function $\mathcal{A}$ depends only on the symmetric momentum variable 
$s_0 = (1/6)\cdot(p_1^2 + p_2^2 + p_3^2)$. 
The model dressing function $\mathcal{A}$ is taken from a DS calculation 
following ref.\ \cite{Eichmann:2014xya}. 
However, here we do not include all of the self-energy contributions which were considered 
in this reference but instead choose a ``ghost-loop-only'' approximation as depicted in 
fig.~\ref{fig:3g_dse}. The resulting dressing function is shown in fig.~\ref{fig:3g_model}. 
Note that these restrictions, in terms of the employed tensor structures, the momentum
dependence and the only kept diagram, are well justified by previous results 
on the three-gluon vertex \cite{Williams:2015cvx, Eichmann:2014xya, Blum:2014gna}
which in turn are substantiated by lattice results \cite{Cucchieri:2008qm} although
in four dimensions they are somewhat inconclusive at lower energies due to the large 
statistical uncertainties. (NB: For a discussion of the technical difficulties to extract
the three-gluon vertex from lattice gauge-field configurations see ref.\ 
\cite{Vujinovic:2018nqc}.)

Here a remark is in order: All Yang-Mills input is taken from DS calculations 
which were originally performed for QCD, {\it i.e.}, for an SU(3) gauge theory. 
Nevertheless, these functions can equally serve as input 
into our SU(2) calculation without any changes due to the choice of 
truncating the DS equations. In the ghost and gluon propagator as 
well as the three-gluon vertex computations specified in ref.\ \cite{Eichmann:2014xya} 
only those diagrams were retained which are proportional to the product $g^2 N_c$, 
with $g$ the gauge coupling and $N_c$ the number of colors. 
Thus one can easily account for the difference in 
the number of colors by changing the renormalization condition
for the running coupling accordingly. The product $g^2 N_c$ remains the same as in 
QCD, and all DS equations are formally remain unchanged. 
This simple trick would have been impossible if, {\it e.g.}, the unquenching effects 
were taken into account for either of the vertex functions. 

\subsection{Numerical results for the fermion propagator and fermion--gauge-boson vertex}

With the Yang-Mills input specified, we briefly comment on the solution method for the
 coupled set of equations for the fermion propagator and fermion--gauge-boson vertex.
We use a fixed-point iteration technique, starting
with an initial guess for the fermion dressing functions and the respective field 
renormalization constant $Z_2$. Note that in the chiral limit a single 
renormalization condition for fermions is sufficient. In addition, this quantity is
ultraviolet finite in the Landau gauge. From $Z_2$ and the ghost propagator input
the fermion--gauge-boson vertex renormalization constant $Z_{1f}$  is determined
from a simple identity
which is valid in the mini-MOM scheme \cite{vonSmekal:2009ae} in Landau gauge:
$Z_{1f} = Z_2/\tilde{Z}_3$. As in this work we will report only on results in the 
chiral limit, 
this fixes all the ingredients needed in the vertex DS equation. Therefore,
 the vertex can be evaluated and back-fed into the quark propagator DS equation
until convergence is reached. All further details, and especially how the internal 
vertex is obtained from tree-level projection of the fully calculated 
vertex, are delegated to Appendix~\ref{app:quark_gluon}. 

\begin{figure*}[t]
\begin{center}
\includegraphics[width = 0.48\textwidth]{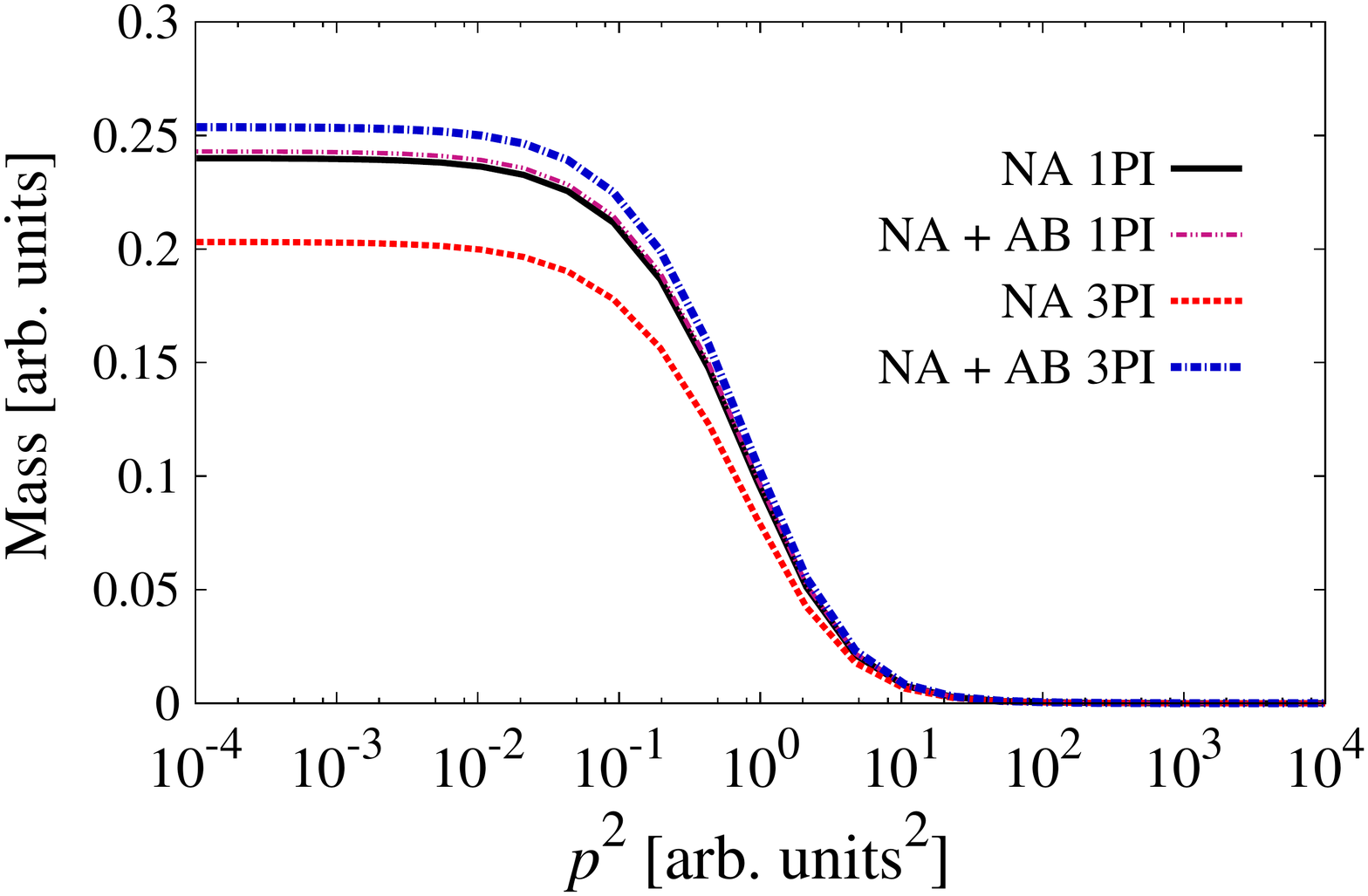}\hspace{0.3cm}
\includegraphics[width = 0.48\textwidth]{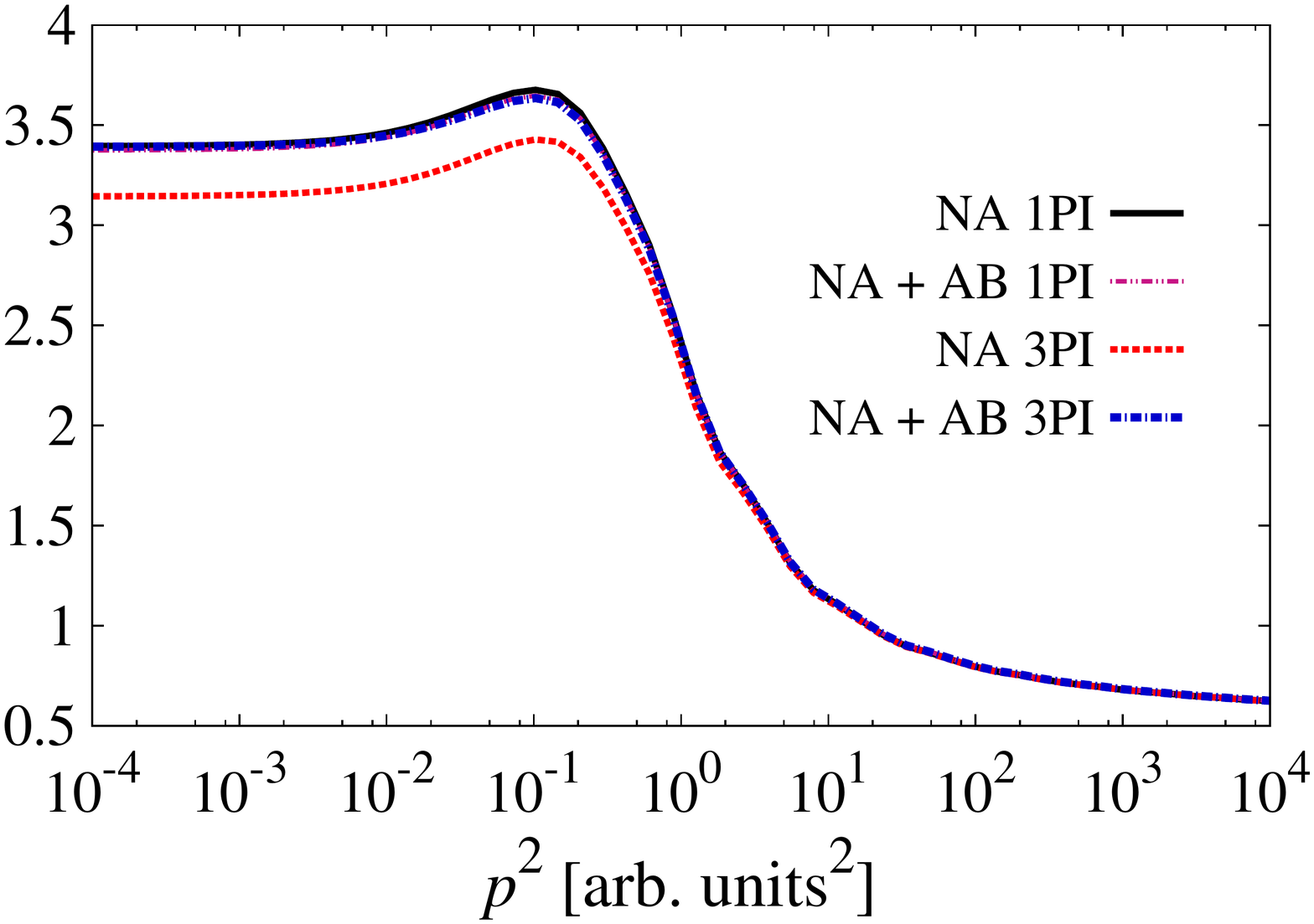}
\caption{The dynamically generated fermion mass function (\textit{left}) and the 
fermion--gauge-boson tree-level tensor structure $T_1(p^2, 2p^2, 3p^2)$ (\textit{right}) 
in different truncations for the fermion--gauge-boson DS equation. 
``NA'' labels the calculation with the non-Abelian diagram only, ``AB'' the 
one with the inclusion of the Abelian one. 
The labels 1PI/3PI are explained in the text and in the caption of 
fig.\ \ref{fig:qg_dse}.}\label{fig:qg_trunc}
\end{center}
\end{figure*}

In fig.\ \ref{fig:qg_trunc} the corresponding results for the fermion mass function 
and the dominant (tree-level) tensor structure of the fermion--gauge-boson vertex
are displayed for several truncations. As noted before, we consider both, the 
1PI- and 3PI-type of DS equations, and we also study a truncation in which 
only the non-Abelian diagram (NA) has been retained in the vertex equation 
in order to probe the relative strengths of various contributions.
One of the first things to note is that in the 1PI-based truncation the influence of 
the Abelian diagram is virtually non-existing compared to the non-Abelian one.
This is in accordance with previous results for the quark-gluon 
\cite{Williams:2009wx} and scalar-gluon \cite{Hopfer:2013via} vertex functions.
However, we note already here that 
despite its negligible effect on these fundamental vertex functions the Abelian 
diagram induces a moderate correction for meson masses in 1PI formulation
as will be seen in the next section.
It would thus not be entirely correct to assume that these diagrams can be 
completely neglected in the 1PI approach, at least when bound state studies 
are concerned. 

By considering the results with the non-Abelian diagram alone one can see that the dressed 
three--gauge-boson vertex has an appreciable impact leading to a significant reduction 
in the dressing functions.
The screening effects of the full gauge-boson three-point correlation
function in the 3PI-based approach are almost canceled exactly by the dressed third 
fermion--gauge-boson vertex in the Abelian diagram. Due to this cancelation 
(which may or may not be coincidental)
the final results are almost identical in the 1PI- and 3PI-based approaches. 

A further test of how close the results for the fermion propagator are in these
two different approaches is provided by its spectral functions. To this end, we
calculate the fermion scalar spectral function by Fourier 
transforming $\sigma_S(p^2) \, = \, Z_f(p^2) M(p^2) \, / \, \bigl( p^2+M^2(p^2) 
\bigr)$:
\begin{equation}\label{eqn:fourier_quark}
\sigma_S(t) = \int d^3 x \int \frac{d^4 p}{(2\pi)^4} e^{ip\cdot x} 
\sigma_S(p^2) \;.                                   
\end{equation}
\begin{figure}[b]
\begin{center}
\includegraphics[width = 0.47\textwidth]{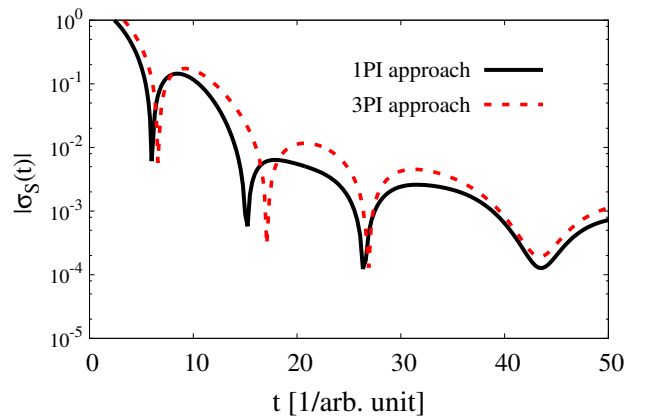}
\caption{Absolute value of the dressing function of Eq.~\eqref{eqn:fourier_quark}, in 1PI and 3PI formulations. Cusps indicate zero crossings in $\sigma_S(t)$.}\label{fig:qrk_posit}
\end{center}
\end{figure}
In fig.~\ref{fig:qrk_posit} the absolute values of $\sigma_S(t)$ in both approaches
are displayed. The cusps in the curves correspond to zero crossings of $\sigma_S(t)$ 
and signal positivity violation for the fermions. First of all, we note again 
the close proximity of the results for the two different approaches.

\begin{figure*}[t]
\begin{center}
\includegraphics[width = 0.9\textwidth]{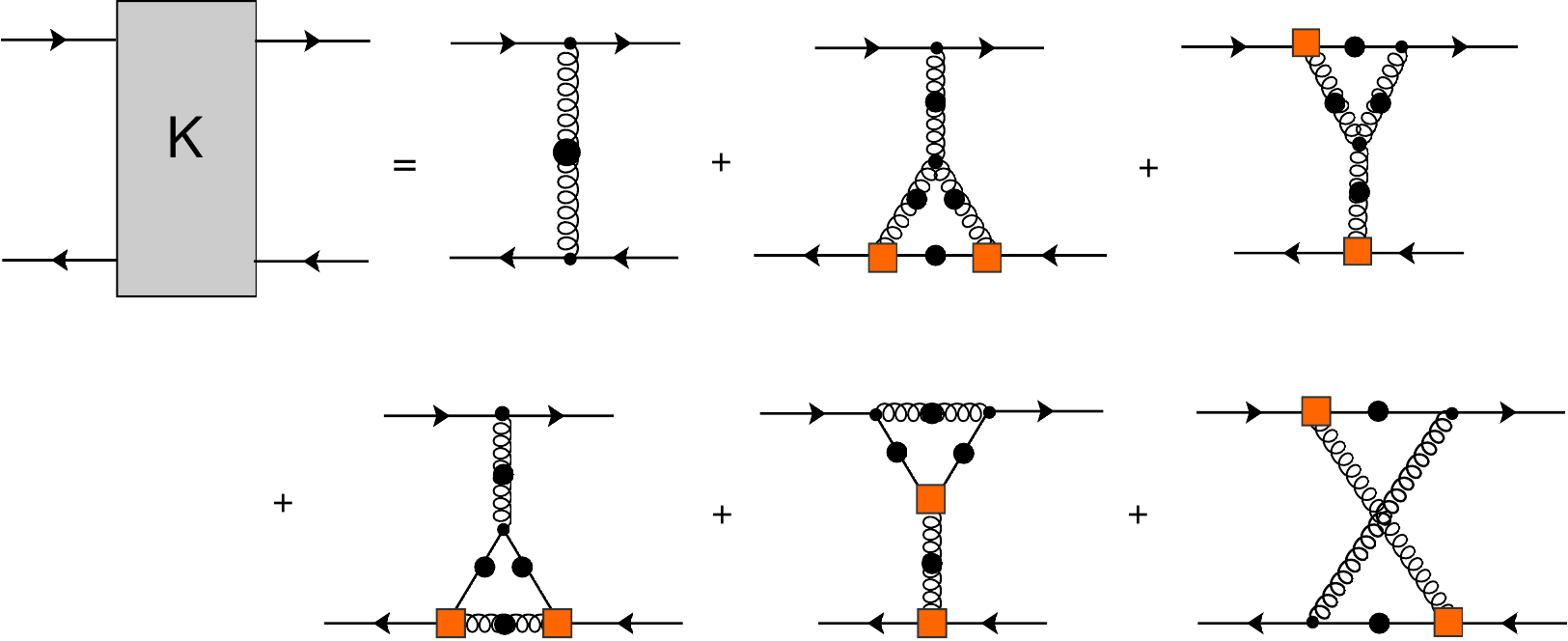}
\caption{A symmetry-preserving truncated BS kernel in agreement with 
the truncated fermion--gauge-boson vertex DS equation  
in fig.~\ref{fig:qg_dse}.}\label{fig:brl_kernel}
\end{center}
\end{figure*}

With respect to the infered positivity violation a remark is in order here.
Patterns similar to the one depicted in fig.~\ref{fig:qrk_posit} were found also in some 
simpler truncation schemes, {\it e.g.}, in the rainbow approximation to the 
quark propagator DS equation, {\it cf.}~ref.~\cite{Alkofer:2003jj} and
references therein. However, the investigation reported in \cite{Alkofer:2003jj}
provided hints that the quark positivity violation within the rainbow approach is merely 
a truncation artifact. The negative norm contributions to $\sigma_S(t)$ vanish upon 
the insertion of some additional tensor structures in the quark-gluon vertex.
Although the here reported results are much more robust the observed positivity violation
might still be an artifact of the approximate treatment of the fermion--gauge-boson 
vertex. We are nevertheless confident that the truncation errors for the fermion propagator 
and the fermion--gauge-boson are small enough to have no substantial impact on the 
bound-state spectra.

To summarize this subsection, it is encouraging that the final results with 
both one-loop diagrams included 
are almost insensitive to whether the 1PI or 3PI framework has been chosen. 
Since the additional vertex dressings in the 3PI version can be seen as an 
approximation to the effective re-summation of certain two-loop terms
this leaves the possibility that the impact of the neglected terms 
is not overwhelmingly large, and that by neglecting them we have
not made an error of qualitative importance. However, it should be pointed out 
that the aforementioned cancelation between the dressed three--gauge-boson 
and fermion--gauge-boson vertex in the 3PI formalism is almost certainly 
restricted to an SU(2) gauge theory. Given the change in color factors such 
a cancelation will not be present to such a high degree in a QCD calculation. 
Noting that the non-Abelian diagram carries a factor $N_c$ and the Abelian
one is suppressed by a factor of $1/N_c$ one can predict an even smaller
impact of the Abelian diagram in QCD. 

%
%*******************************************************************
% complete BS equation
%*******************************************************************
%
%
\subsection{Truncations for the kernel of the bound-state equations}
\label{sec:kernel}

Besides the propagators and vertex functions discussed above
the most important ingredient into the reported calculation is 
the symmetry-preserving BS kernel as obtained with the cutting technique of 
eq.~\eqref{eqn:cutting}. For the truncated fermion--gauge-boson vertex DS equation  
shown in fig.~\ref{fig:qg_dse} the corresponding truncated kernel is displayed in 
fig.~\ref{fig:brl_kernel}. In the 3PI-based truncation 
there are additional dressings for the three--gauge-boson and 
fermion--gauge-boson vertices. As it is straightforward to implement them 
we are refraining from showing them explicitly.

%
%*******************************************************************
% SU(2) ground state spectrum
%*******************************************************************
%
%
\section{Ground state mesons}\label{sec:mesons}

\begin{table*}[t]
\begin{center}
\caption{Ground state masses in various truncations of the employed DS and BS 
equations in internal units:
 ``NA'' stands for the non-Abelian diagram only, ``AB'' for including the Abelian one. 
The given errors are purely numerical and are estimated within the employed 
extrapolation procedure, see text for details.}
\label{tab:arb_units_masses}
\bigskip
\setlength{\tabcolsep}{14.5pt} % to change column spacing
\begin{tabular}{c|c|c|c|c}
 \hline\hline\noalign{\smallskip} % top of table
$m_{J^{PC}}$           &    NA, 1PI   &  NA + AB, 1PI   &     NA, 3PI    &  NA + AB, 3PI \\
\noalign{\smallskip}\hline\noalign{\smallskip}
$m_{0^{-+}}$%\, / f_{PS}$    
&    0         &    $ 0      $   &   $0   $       &    0  \\   
$m_{0^{++}}$%\, / f_{PS}$    
&   385(8)     &    $ 358(7) $   &   $335(7) $    &    356(7)  \\
\noalign{\smallskip}\hline\noalign{\smallskip}
$m_{1^{--}}$%\, / f_{PS}$   
 &   628(13)    &    $ 583(12) $  &   $ 597(12) $  &    567(11)  \\
$m_{1^{++}}$%\, / f_{PS}$    
&   794(15)    &    $ 775(14) $  &   $ 778(14) $  &    760(14)  \\
\noalign{\smallskip}\hline\noalign{\smallskip}
$f_\pi$     &     68       &      72         &      62        &     70  \\ 
\noalign{\smallskip}\hline\hline % bottom of table
\end{tabular}
\end{center}
\end{table*}

Although our study is motivated to a large part by exploring possibilities for 
a theory of Beyond-the-Standard-Model physics we will mostly not concern ourselves 
with the composite Higgs and/or technicolor aspects of the considered gauge theory. 
The aim here is to obtain the ground state spectrum of  $J \leq 1$ bound states 
and compare them with lattice results of refs.~\cite{Arthur:2016dir,Arthur:2016ozw}.
In addition, to allow for a further development of truncations, we are interested
how the calculated ground state spectrum is influenced by the different contributions
in the DS equations, and hereby especially in the DS equation for the fermion--gauge-boson
vertex. Of course, such tests can be performed completely with arbitrary internal units only.
Therefore, we will choose a scale more for the matter of convenience than for
necessity. The quantity which is fixed in scale-setting is the Higgs vacuum
expectation value, $v_{ew}=$ 246 GeV. On a purely formal level, this quantity is
identical to the ``pion'' decay constant, {\it i.e.}, the decay constant of the 
pseudoscalar Goldstone fields. It is then calculated via the identity~\cite{Tandy:1997qf}:  
\begin{align}\label{eqn:piondecay}
f_{PS} = \frac{Z_2 N_c}{\sqrt{2}P^2}\text{tr}\int_k\Gamma_{PS}(k,-P)S(k_+)
\gamma_5\slashed{P}S(k_-),
\end{align}
with $k_\pm = k \pm P/2$, and $\Gamma_\pi$ being the Goldstone boson BS amplitude, 
normalised with the Nakanishi condition \cite{Nakanishi:1965zz}. 
(NB: In the above relation we employ conventions for which $f_\pi = 93$ MeV in QCD.)

The bound state masses in various truncations for the employed DS and BS 
equations are shown in table~\ref{tab:arb_units_masses}. 
As only the chiral limit is considered the pseudoscalar, {\it i.e.}, Goldstone boson,
states are strictly massless.
The results are displayed in arbitrary units in order to separate the direct 
influence of the truncations from the scale setting procedure.
We provide the numerical value of $f_\pi$ in the last row of the 
respective columns to allow for transformation to physical scales.
The errors on the masses are purely numerical and stem from the chosen way
to extract the masses. In Euclidean field theory, the bound state on-shell 
condition $P^2 = -M^2$ (with $M$ being the bound state mass) implies
working with complex-valued total momentum $P$.
However, there are ways to reliably estimate some of the hadronic observables by
extrapolating from the region of spacelike $P^2$, see, {\it e.g.}, 
refs~\cite{Bhagwat:2007rj,Dorkin:2013rsa,Tripolt:2018xeo} and references therein.
Here
we employ the inverse vertex extrapolation technique, which is explained in detail in 
\cite{Bhagwat:2007rj}. Errors for some of the meson masses in 
table~\ref{tab:arb_units_masses} come from this approximate
treatment.~The efficiency of the method was thoroughly tested in 
\cite{Vujinovic:2014ioa}, and it was found to be very reliable, at least for relatively light 
$J\leq 1$ mesons which we consider here.~Note that, in
the chiral limit, the Goldstone boson decay constant $f_{PS}$ is one of the very 
few quantities which can be calculated exactly purely from spacelike momenta,
or more precisely for $P^2\ge0$. 
For this reason, the scale setting procedure does not
introduce any additional uncertainties. 

Let us take the results in the ``NA, 1PI'' column of Table \ref{tab:arb_units_masses} 
as the point of reference. Comparison with other truncations shows that all the modifications 
(addition of Abelian loops, the 3PI
vertex dressings), induce moderate corrections. In terms of relative mass differences, the 
$0^{++}$ channel seems to be most susceptible to various approximations, whereas the 
vector mesons (especially the axial one) are
somewhat robust in this regard.
The Abelian diagram induce modest relative changes to the bound state masses, ranging 
from five to ten percent across different channels. 
Note that not only for QCD but also for all larger gauge groups 
(which have also been investigated with respect to 
Beyond-the-Standard-Model physics, see, {\it e.g.}, ref.~\cite{Llanes-Estrada:2018azk} 
and references therein) the impact of these diagrams would be further 
suppressed by group-theoretical factors.

It is interesting to note that the mass of the scalar in this calculation seems to be
only mildly influenced by BRL effects. In the 1PI-based approach (including the Abelian
diagram) one has $m_{0^{++}}\, / f_{PS} = 5.0 \pm 0.1$ and for the 3PI one
$m_{0^{++}}\, / f_{PS} = 5.1 \pm 0.1$, {\it i.e.}, both values are 
very close to the RL result \cite{Vujinovic:2014ioa}. This is in contrast
to the calculation of the scalar meson mass in ref.~\cite{Williams:2015cvx}:
The RL value $m_{0^{++}}\, / f_{\pi} = 6.96$ is significantly lower than the 
obtained value in the much more sophisticated 3PI truncation employed there,
$m_{0^{++}}\, / f_{\pi} = 10.5 \pm 1.0$. Therefore, this comparison provides evidence
that (i) the bound-state masses for an SU(3) gauge theory are much larger than for a 
SU(2) one
(NB: This result is in agreement with the analysis of dynamically generated fermion masses in
ref.\ \cite{Llanes-Estrada:2018azk}),  
and (ii) BRL effects in this channel are more pronounced for SU(3) than SU(2). Having 
a look at the vector and axialvector channels these differences seems to be much smaller.
The QCD study of ref.~\cite{Williams:2015cvx} provides for the 3PI--3-loop truncation
$m_{1^{--}}\, / f_{\pi} = 7.0$ whereas we obtain $m_{1^{--}}\, / f_{PS} = 8.1 \pm 0.2$.
As usual BRL effects are small for the vector channel, {\it cf.}, the discussion in the 
review \cite{Eichmann:2016yit} where the presence / absence of BRL corrections is related 
to importance / insignificance of spin-flip type interactions among the constituents. 
For the axialvector channel BRL effects are significant, and the SU(3) versus SU(2) 
comparison points towards an even smaller difference or maybe even
almost no difference within the numerical accuracy, 
$m_{1^{++}}\, / f_{\pi} = 12.4\pm 1.0$ versus $m_{1^{++}}\, / f_{PS} = 10.9 \pm 0.2$

\begin{table*}[t]
\begin{center}
\caption{Light meson masses in various truncations of the quark-gluon DSE and the meson BSE, compared with the 
lattice data. All results are in units of TeV. The $0^{++}$ mass is for an isoscalar: the corresponding lattice 
input is forthcoming.}\label{tab:phys_units_masses}
\setlength{\tabcolsep}{14.5pt} % to change column spacing
\begin{tabular}{c|c|c|c|c|c}
\hline\hline\noalign{\smallskip} % top of table
$J^{PC}$    &    NA, 1PI    &  NA + AB, 1PI    &     NA, 3PI     &  NA + AB, 3PI  & Lattice \cite{Arthur:2016dir,Arthur:2016ozw}\\
\noalign{\smallskip}\hline\noalign{\smallskip}
$0^{-+}$    &    0          &    $ 0       $   &   $0    $       &    0           &   --\\   
$0^{++}$    &   1.39(3)     &   $ 1.22(2) $    &   $1.33(3) $    &    1.25(2)     &   
$4.7 \pm 2.7 $\\
\noalign{\smallskip}\hline\noalign{\smallskip}
$1^{--}$    &   2.27(5)    &    $ 2.00(4) $  &   $ 2.37(5) $  &    1.99(4)        &  $ 3.2 \pm 0.5 $ \\
$1^{++}$    &   2.87(5)    &    $ 2.65(5) $  &   $ 3.09(6) $  &    2.67(5)        &  $ 3.6 \pm 0.9 $ \\
\noalign{\smallskip}\hline\hline % bottom of table
\end{tabular}
\end{center}
\end{table*}

The differences discussed above are important for future investigations which aim 
at a good numerical precision. Certainly, more work is needed to judge whether a level
of ``apparent convergence'' has been reached already. 
From a qualitative perspective, however, one can note already some interesting trends.
The ground state spectrum of the theory investigated here has also been studied 
on the lattice \cite{Arthur:2016dir,Arthur:2016ozw}, but as it is shown in 
table~\ref{tab:phys_units_masses}, the corresponding results have relatively 
large uncertainties, definitely larger ones than in between our different truncations.
It is interesting to note that the masses as obtained on the lattice seem to be 
systematically larger than our results. As a rule of thumb one may summarise the 
comparison by the statement that the masses from the DS equations are located at the 
lower end of the 1-$\sigma$-band of the lattice results.

An inspection of table~\ref{tab:arb_units_masses} reveals that all of the 
improvements of the simplest ``1PI, NA'' scenario which we have considered supress
the masses further. Taken together with the influence of pion back-reaction, 
as investigated in \cite{Williams:2009wx}, one is drawn towards the conclusion 
that all of the one-loop corrections to the  fermion--gauge-boson vertex beyond
the simplest non-Abelian treatment invariably have a screening effect on the  results
for bound states, seemingly pushing them away from experimental data in QCD, or 
central lattice estimates in an SU(2) technicolor theory. This motivates than 
a fully self-consistent calculation for different gauge groups, especially 
because the meson mass results of ref.\ \cite{Williams:2015cvx} already show quite 
some improvement in this regard. 

%
%
%***********************************************************************
% Conclusions
%***********************************************************************
%
%
\section{Conclusions}\label{sec:conclude}

Building on the established knowledge of correlation functions in gauge theories
in Landau gauge we have studied the effects of various truncations 
of the DS equations for the fermion--gauge-boson vertex, the fermion propagator,
and the ground state spectrum in an SU(2) gauge theory.  For both, the fundamental vertex 
functions and the bound state masses, we found relatively mild changes in results for 
different truncations. For the masses they were on the order of roughly five to ten 
percent. Whereas the recent respective lattice results \cite{Arthur:2016dir,Arthur:2016ozw} 
seem to indicate a significant difference to results in an SU(3) gauge theory we obtained
masses which were consistently lower than the central values of the lattice results but are
nevertheless in agreement with them at an approximately 1-$\sigma$-level. Therefore 
differences to the SU(3) meson spectrum remain, especially for the scalar, but they 
are not as pronounced as indicated by the (central values of the) lattice results.

It remains to be seen if the methods outlined here can lead to certain improvements 
when applied to the baryon sector of QCD, for instance in the description of the 
nucleons' negative parity partner~\cite{Sanchis-Alepuz:2015qra}.
Also, the fact that even in our simplified framework the influence of the 
Abelian diagram in the  fermion--gauge-boson equation was found to be modest
but still noticeable, is suggestive that in a self-consistent calculation these 
terms might induce potentially significant corrections because in the BS kernel
there are then four fully dressed fermion--gauge-boson vertices 
\cite{Sanchis-Alepuz:2015tha,Williams:2015cvx}. 

In summary, the here presented investigation has provided one further step towards
a fully self-consistent treatment of gauge-invariant bound states
in gauge theories with a sufficiently sophisticated truncation scheme
of DS and BS equations. 
Establishing such a scheme will first of all provide more insight into the 
binding mechanisms for highly-relativistic bound states. For
theories with a walking behaviour of the coupling (which implies that one deals 
with a multi-scale problem) functional methods based on continuum quantum field 
theories may offer even a higher precision than lattice calculations.

\bigskip

\section*{Acknowledgments}

MV gratefully acknowledges financial support by the University of Graz and by the 
Austrian science fund FWF via the Doctoral Program W1203 and the Schr\"odinger
grant J3854-N36.

We thank H\`elios Sanchis-Alepuz for his critical reading of this manuscript and
his comments.

\newpage

%
%
%***********************************************************************
% Appendices
%***********************************************************************
%
%
\appendix

\section{Numerical method for the calculation of the 
fermion--gauge-boson vertex}\label{app:quark_gluon} 

In the treatment of the fermion--gauge-boson vertex we implement a scheme
described in ref.~\cite{Williams:2014iea}. The main idea is to 
use on the r.h.s.\ of the vertex equation a projection of  
the vertex on its tree-level tensor structure and therefore significantly
simplified internal vertices. This projection
is achieved by constructing an effective
dressing function $\lambda(k^2)$, {\it cf.} eq.~\eqref{eqn:rainbow}. 
To be explicit, the following parameterization is used:
\begin{align}
\lambda(k^2) & = h Z_{1f}\biggl\{\frac{L(M_0)}{1+y} \nonumber \\ 
             & + \frac{1}{1+z}\left[\frac{4\pi}{\beta_0\alpha_\mu}\left(\frac{1}{\log(x)} - \frac{1}{1-x}\right)\right]^{18/44}\biggr\}\;,             
\end{align}
with $h= 2.302,~x = k^2/0.6,~y = k^2/0.34,~z = k^2/0.33,~\beta_0 = 11N_c/3$, 
and $\alpha_\mu = 1.114$ the renormalized coupling at a scale $\mu = 3$ 
(in arbitrary units). The infrared enhancement $L(M_0)$ depends on the quark
mass at zero momentum ($M_0=M(p^2)=0$) and is parameterized as a ratio
of polynomials:
\begin{equation}\label{eqn:ir_enh}
L(M_0) = \frac{a + bM_0 + cM^2_0}{M_0 + dM_0^2} \;. 
\end{equation}
The coefficients $a,b,c$ and $d$ are determined such that the total model 
dressing $\lambda(k^2)$ fits reasonably precisely the tree-level projection of the 
full calculated fermion--gauge-boson vertex. These parameters are given in
table~\ref{tab:qg_model} for the truncations considered in this work.  

\begin{table}[ht]
\begin{center}
\caption{Coefficients of Eq.~\eqref{eqn:ir_enh} in various approximations for the 
fermion--gauge-boson vertex. (NB: The ``NA + AB, 1PI'' parameters have the same
numerical values as ones of ``NA, 1PI''.)}
\label{tab:qg_model}
\setlength{\tabcolsep}{9.5pt} % to change column spacing
\begin{tabular}{c|c|c|c}
\hline\hline\noalign{\smallskip} % top of table
Coeff.   &    NA, 1PI   &  NA, 3PI   &   NA + AB, 3PI    \\
\noalign{\smallskip}\hline\noalign{\smallskip}
$ a $    &   0.244      &  -0.014    &   0.342      \\   
$ b $    &   1.788      &   3.20     &   1.037    \\
\noalign{\smallskip}\hline\noalign{\smallskip}
$ c $    &   -0.198     &   0.842    &  -0.812   \\
$ d $    &   0.293      &   1.928    &  -0.621    \\
\noalign{\smallskip}\hline\hline % bottom of table
\end{tabular}
\end{center}
\end{table}

For completeness we provide the covariant tensor decomposition for the full 
calculated vertex $\Gamma^\mu(p_1, p_2)$, with $p_1$ and $p_2$ denoting, respectively, 
the incoming and outgoing fermion momenta.
Defining the relative momentum $l = (p_1 + p_2)/2$, and 
the outgoing gluon momentum $k = p_2 - p_1$, we use the orthonormal combinations:
\begin{align}
& t^\mu = \hat{k}^\mu \;, \nonumber\\
& s^\mu = \hat{h}^\mu \; {\mathrm {with}} \; h^\mu = T^{(t)}_{\alpha\beta}l^\beta \;, \nonumber \\
& \gamma^\mu_{TT} = {T^{(t)}}^{\mu\alpha}T^{(s)}_{~~~\alpha\nu} \, \gamma^\nu 
\, = \, \gamma^\mu - \slashed{t}t^\mu - \slashed{s}s^\mu \;,
\end{align}
with the hat denoting normalization of the corresponding four-vector. 
Any components proportional to $t^\mu$ will be projected out in Landau gauge, 
leaving eight purely transverse tensor components. Their basis is chosen to be
\begin{equation}
(\gamma_{TT}^\mu, s^\mu)\times(\mathbf{1},\slashed{s},\slashed{t},\slashed{s}\slashed{t}) \;.
\end{equation}
Here, $\mathbf{1}$ stands for a Dirac unity matrix.
The $T_1$ dressing functions, plotted in the right panel of fig.~\ref{fig:qg_trunc}, 
is the prefactor of the tensor $\gamma_{TT}$.

\newpage
%
%
%*******************************************************************
% Bibliography
%*******************************************************************
%
%
\begin{widetext}

\end{widetext}
\end{document}